# Superconductivity in $Ba_{1-x}K_xTi_2Sb_2O$ ($0 \leq x \leq 1$) controlled by charge doping

Ursula Pachmayr and Dirk Johrendt[*]

Department Chemie, Ludwig-Maximilians-Universität München

Abstract

The solid solution of antimonide-oxides $Ba_{1-x}K_xTi_2Sb_2O$ ($0 \leq x \leq 1$) has been synthesized by solid-state reactions and characterized by X-ray powder diffraction ($CeCr_2Si_2C$-type structure; $P4/mmm$, $Z = 1$). The crystal structure consists of $Ti_2Sb_2O$-layers that are stacked with layers of barium atoms along the $c$-axis. $BaTi_2Sb_2O$ is a known superconductor with a critical temperature ($T_c$) of 1.2 K. Substitution of barium through potassium raises $T_c$ up to 6.1 K at 12 % potassium, while no superconductivity emerges with concentrations higher than 20 %. Anomalies in electrical transport and magnetic susceptibility indicate charge density wave (CDW) instabilities. The CDW transition temperatures ($T_a$) decrease from 50 K in the parent compound to 28 K at 10 % potassium substitution. No CDW transition was detected at higher concentrations, and no evidence for a reduction of the lattice symmetry below $T_a$ was found. The lattice parameters vary linearly while the unit cell volume increases with higher potassium concentrations. The phase diagrams $T_c(x)$ and $T_a(x)$ of $Ba_{1-x}K_xTi_2Sb_2O$ are remarkably similar to the known series $Ba_{1-x}Na_xTi_2Sb_2O$ ($0 \leq x \leq 0.33$) in spite of the reverse volume effect. From this we conclude that the charge and not the volume determines the phase diagrams of these superconducting antimonide-oxides.

[*]Prof. Dr. Dirk Johrendt

Department Chemie, Ludwig-Maximilians-Universität München

Butenandtstrasse 5-13 (Haus D), 81377 München, Germany

Phone +49 (0) 89-2180-77430; Fax +49 (0) 89-2180-77430

Email: dirk.johrendt@cup.uni-muenchen.de

# 1 Introduction

Transition metal pnictides with layered crystal structures attract considerable interest in particular after the discovery of high-$T_c$ superconductivity in iron-arsenides [1-5]. It is generally accepted that superconductivity in iron-arsenides emerges in the anti-fluorite like FeAs-layers during suppression of antiferromagnetic spin-density wave (SDW) ordering in parent compounds like LaOFeAs or BaFe$_2$As$_2$, either by chemical doping or by applying pressure. Today it is considered that the formation of the Cooper pairs in iron-based superconductors is unconventional, and mediated by magnetic spin fluctuations [6,7]. The proximity of superconductivity to magnetic (possibly nematic [8]) SDW ordering and lattice instabilities became evident by the competition of these order parameters, which partially co-exist microscopically in the underdoped areas of the phase diagrams [9,10].

Charge-density waves (CDW) represent another well-known type of electronic order [11], which has also been discussed to play a role in the pairing mechanism of certain superconductors, among them transition-metal dichalcogenides like TaS$_2$ [12]. Recently a CDW instability which competes with superconductivity has been observed in copper-oxides by resonant soft X-ray scattering [13,14].

A long known pnictide-oxide with CDW instability is the antimonide-oxide Na$_2$Ti$_2$Sb$_2$O with a layered crystal structure that consists of [Ti$_2$Sb$_2$O]$^{2-}$-slabs separated by double layers of sodium ions [15,16]. Strong anomalies in the resistivity and magnetic susceptibility as well as a structural distortion occur at 120 K ($T_a$), while no superconductivity has been detected. Identical [Ti$_2$Sb$_2$O]$^{2-}$-slabs are present in the closely related compound BaTi$_2$Sb$_2$O, which exhibits analogue but much weaker anomalies near ≈50 K and superconductivity at 1.2 K [17]. These anomalies in the resistivity, susceptibility and specific heat have been ascribed to CDW ordering, supported by DFT calculations of BaTi$_2$Sb$_2$O that revealed Fermi-surface nesting [18]. Sodium doping increases the superconducting critical temperature from 1.2 to 5.5 K in Ba$_{1-x}$Na$_x$Ti$_2$Sb$_2$O at $x \approx 0.25$, while the CDW ordering temperature decreases from 54 K to 38 K [19]. It has been argued whether this is a new example of competing orders in the proximity of superconductivity [17,19], and even a scenario for unconventional pairing in Ba$_{1-x}$Na$_x$Ti$_2$Sb$_2$O has been suggested [18]. However, several recent papers conclude that superconductivity in these materials is conventional in nature, and mediated by electron-phonon coupling [20-22]. Myon spin rotation (μSR) experiments additionally showed that the anomaly in the magnetic susceptibility is no magnetic order, thus the transition is very probably the formation of a CDW and not a SDW [23].

Even though superconductivity in doped $BaTi_2Sb_2O$ compounds turned out conventional, the relation with the CDW ordering is still unclear. In this study we present the synthesis, structure and properties of the complete solid solution $Ba_{1-x}K_xTi_2Sb_2O$ ($0 \leq x \leq 1$). Potassium and barium ions are similar in size, thus we are able to study how charge variations act on the properties without volume change. We show that the phase diagram obtained by potassium doping is remarkably similar to that by sodium doping in spite of the reverse volume effect.

## 2    Experimental

Polycrystalline samples of $Ba_{1-x}K_xTi_2Sb_2O$ ($x$ = 0, 0.04, 0.05, 0.1, 0.12, 0.15, 0.2, 0.3, 0.4, 0.5, 0.7, 1.0) were synthesized by solid-state reactions using the starting materials Ba (99.99%), $BaO_2$ (99.99%), K (99.95%), Sb (99.999%) and Ti (99.5%) for $x \leq 0.5$ and Ba, K, Sb, Ti and $TiO_2$ (99.995%) for $x > 0.5$. Stoichiometric mixtures of the starting materials were sealed in welded niobium tubes within silica ampoules under purified argon atmosphere and slowly heated to 600 °C, holding for 15 h before cooling to room temperature. After the first reactions, the products were ground, pelletized and sintered at 900 °C for 3 days followed by a rather slow cooling to 200 °C. Additional grinding and sintering at 900 °C for another 3 days was performed to ensure phase homogeneity. For the sample $x$ = 0.7 the synthesis temperature was reduced to 750 °C, whereas $KTi_2Sb_2O$ was synthesized at 700 °C after a first temperature step at 550 °C. The air- and moisture-sensitive polycrystalline products had a dark gray color. All experimental handling was performed in a purified argon atmosphere glove box.

Powder X-ray diffraction was carried out at room temperature using a Huber G670 diffractometer with $Cu_{K\alpha 1}$ radiation ($\lambda$ = 1.5405 Å) and Ge-111 monochromator. For low temperature powder X-ray investigations a Huber G670 diffractometer ($Co_{K\alpha 1}$ radiation, $\lambda$ = 1.7902 Å, Ge-111 monochromator) equipped with a close-cycle He-cryostat was employed. The detailed structural parameters were obtained by Rietveld refinement using the software package TOPAS [24]. The Ba : K ratios determined by refining occupation parameters are in good agreement with the ones obtained by energy dispersive X-ray (EDX) analysis on a scanning electron microscope Jeol JSM-6500 FE SEM. For investigation of the magnetic properties a Quantum Design MPMS-XL5 SQUID magnetometer was used, whereas superconductivity was examined in ac-susceptibility measurements.

## 3 Results and discussion

### 3.1 Crystal structure and lattice parameters

Phase purities of the obtained polycrystalline $Ba_{1-x}K_xTi_2Sb_2O$ samples ($x$ = 0, 0.04, 0.05, 0.1, 0.12, 0.15, 0.2, 0.3, 0.4, 0.5, 0.7, 1.0) were checked by Rietveld-analysis of the X-ray powder patterns. Figure 1 shows the generic pattern of $Ba_{0.5}K_{0.5}Ti_2Sb_2O$. No impurities could be detected within the experimental limitations (~1% of a crystalline phase). The lattice parameters of the $BaTi_2Sb_2O$ parent compound ($a$ = 411.1(1) pm, $c$ = 807.0(1) pm) and the atomic position of the antimony atom agree with the data given by Yajima *et al*. [17]. The crystal structure of $BaTi_2Sb_2O$ has been assigned to the $CeCr_2Si_2C$-type [19]. However, the silicide-carbide exhibits strong bonds between the silicon atoms which form $Si_2$-dimers, while no such bonds exist in $BaTi_2Sb_2O$, where the Sb-Sb distance is above 400 pm. Therefore the structures of $CeCr_2Si_2C$ and $BaTi_2Sb_2O$ are not isotypic, but isopointal [25].

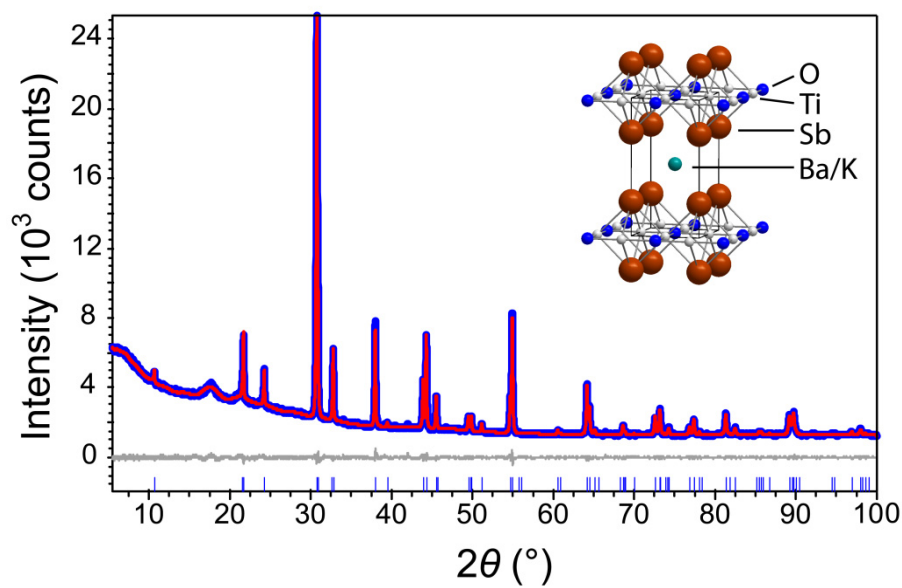

Figure 1: X-ray powder pattern of $Ba_{0.5}K_{0.5}Ti_2Sb_2O$ (blue) with Rietveld-fit (red) and difference plot (grey). Insert: Crystal structure of $Ba_{1-x}K_xTi_2Sb_2O$.

$Ba_{1-x}K_xTi_2Sb_2O$ exhibits a systematic contraction of the *a*-axis, whereas the *c*-axis of the unit cell gradually increases with increasing doping level *x* (Figure 2a). In so doing, the *a*-axis parameter exhibits linear behavior fulfilling Vegard's law. This is broadly similar for the *c*-axis parameter except for the end compound $KTi_2Sb_2O$ featuring a higher value than expected from the linear progression. Concomitant with increasing potassium doping, the Coulomb attraction between $(Ba,K)^{(2-x)+}$ and $Sb^{3-}$ is reduced leading to an elongated distance of Ba/K and Sb, and consequently to an enlargement of the *c*-axis. Additionally, geometrical effects resulting from the larger radius of the $K^+$ ions ($r_{K^+}$ = 151 pm, $r_{Ba^{2+}}$ = 142 pm [26]) together with the shortening of the *a*-axis enhance this enlargement.

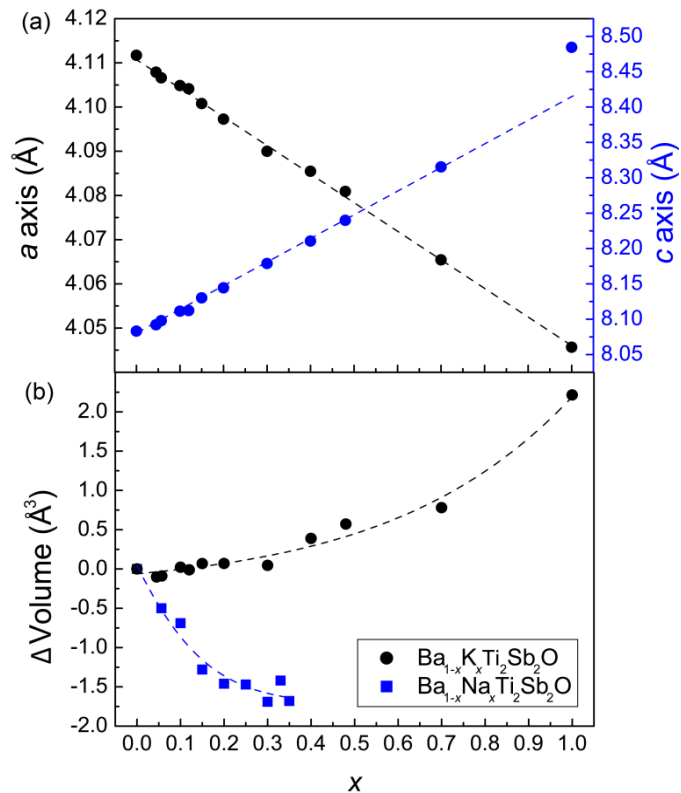

Figure 2: (a) Lattice parameters of $Ba_{1-x}K_xTi_2Sb_2O$. (b) Relative changes of the unit cell volumes in $Ba_{1-x}K_xTi_2Sb_2O$ compared with $Ba_{1-x}Na_xTi_2Sb_2O$ (data of the sodium compound are taken from [19]). Dashed lines are guides to the eye.

As the elongation of the unit cell in *c*-axis direction is larger compared to the contraction in *a*-axis direction (compare Figure 2a) an alteration in the volume of the unit cell can be expected. In Figure 2b the development of the unit cell volume with increasing doping level *x* is depicted. For small doping levels $0 \leq x \leq 0.3$ the volume stays almost constant before it increases continuously at higher doping levels. Consequently, at small doping levels the increase in lattice parameter *c* seems to be right balanced by the reduction in lattice parameter *a*, the latter contributing in the power of two (tetragonal cell). Note that the doping level of 30 % constituting the begin of continuous volume increase coincides with the maximum doping level realized in the known other hole-doped compounds $Ba_{1-x}Na_xTi_2Sb_2O$ and $BaTi_2(Sb_{1-x}Sn_x)_2O$ synthesized so far [19,27]. The comparison with the sodium-doped materials is in particular interesting. The unit cell volume of $Ba_{1-x}Na_xTi_2Sb_2O$ decreases with increasing sodium concentration, which is mainly the consequence of the smaller increase of the *c*-axis parameter. However, data are only available for $0 \leq x \leq 0.35$ [19].

3.2 Superconductivity

Measurements of the ac-susceptibilities revealed superconductivity in $Ba_{1-x}K_xTi_2Sb_2O$ with *x* = 0.04-0.15 (Figure 3). The maximum $T_c$ of about 6.1 K is obtained at the *x* = 0.12 doping level, the transition results in a throughout superconducting sample. This value constitutes not only the maximum $T_c$ in the system $Ba_{1-x}K_xTi_2Sb_2O$ but in the complete $ATi_2Pn_2O$ family so far (*A* = $Na_2$, Ba, $(SrF)_2$, $(SmO)_2$; *Pn* = As, Sb, Bi) [16,17,28]. The superconducting transitions for the *x* = 0.04, 0.05 and *x* = 0.1 samples are steep, and bulk superconductivity militating for the high quality of the samples is achieved. The transition of the *x* = 0.12 sample is slightly broadened, but complete shielding is nevertheless achieved. The fact that the measurement indicates a volume susceptibility $4\pi\chi_V$' of more than 100 % in those samples could be ascribed to various effects such as an inaccuracy in the assumed density of the sample, grain boundary effects, as well as the influence of demagnetization and particle size.

For a doping level of 15 % the superconducting volume fraction is decreased to about 40 % at 3.5 K, whereby it cannot be considered as bulk superconductivity any longer for doping levels of 20 % and higher. Concomitant with the decrease in superconducting volume fraction the transition is obviously broadened. Normally such a broadening in the superconducting transition of doped compounds might be ascribed to inhomogeneous doping distribution.

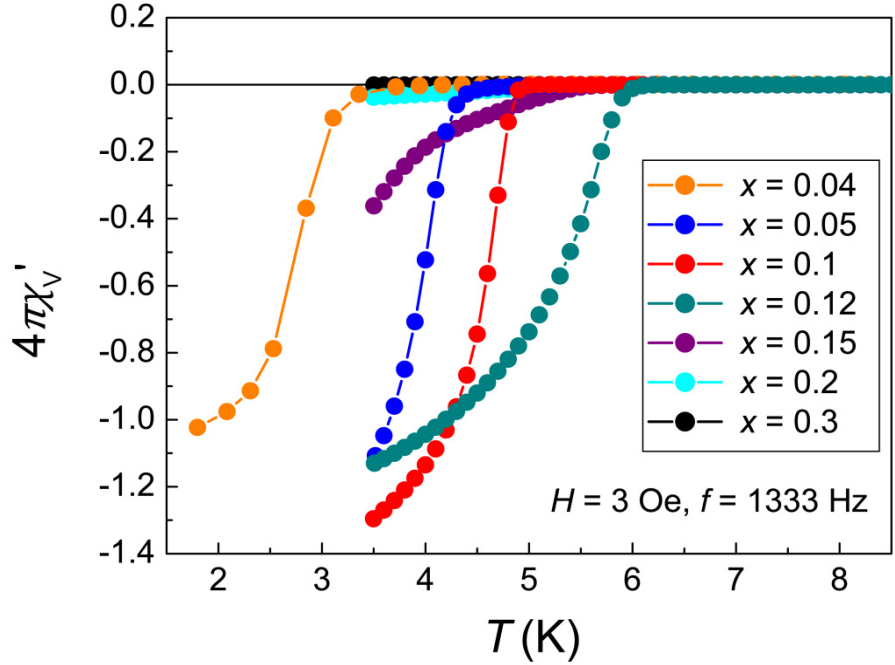

Figure 3: Low-temperature ac-susceptibility of $Ba_{1-x}K_xTi_2Sb_2O$

### 3.3 Magnetic susceptibility

$BaTi_2Sb_2O$ exhibits a magnetic susceptibility anomaly at about 50 K ($T_a$) accompanied by weak features in the resistivity, specific heat, and lattice parameters [17]. The nature of this transition has initially been ascribed to either a charge-density wave (CDW) or spin-density wave (SDW) instability in analogy to the closely related $Na_2Ti_2Sb_2O$ [15]. Indeed a recent DFT study of $BaTi_2Sb_2O$ revealed at least partial Fermi-surface nesting associated with a magnetic instability [18]. Calculated phonon dispersions suggested a weak lattice anomaly and electron-phonon coupling large enough to explain the superconductivity [21]. However, a very recent μSR study of $Ba_{1-x}K_xTi_2Sb_2O$ ruled out the presence of magnetic ordering below $T_a$, which means that the transition is CDW and not SDW [23].

Figure 4 shows the magnetic susceptibilities of $Ba_{1-x}K_xTi_2Sb_2O$ measured at 20 kOe. Correspondingly to the known characteristics of the undoped compound $BaTi_2Sb_2O$ [17], the slightly potassium-doped sample ($x = 0.04$) shows an anomaly in the magnetic susceptibility. This anomaly is shifted from 50 K in the parent compound towards a lower temperature of about 33 K. This shift of the transition to lower temperatures is accompanied with an increase of $T_c$ from 1.2 K [17] to 4.3 K being qualitatively consistent with the results obtained for $Ba_{1-x}Na_xTi_2Sb_2O$ and $BaTi_2(Sb_{1-x}Sn_x)_2O$ [19,27]. With increasing doping level $x$ the anomaly is further shifted towards lower temperatures, becoming increasingly less distinctive. Whereas a determination of $T_a$ can be still accomplished for a doping level of 10 %, no anomaly is discernible at higher doping levels.

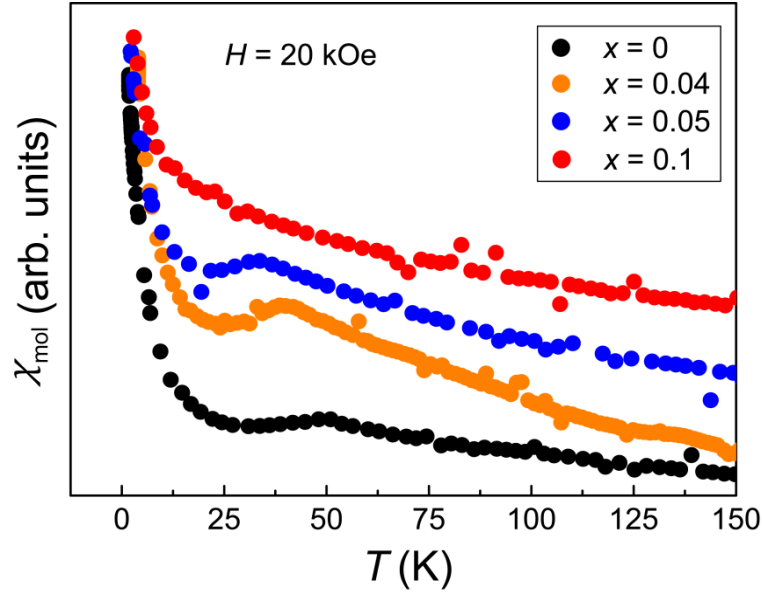

Figure 4: Magnetic susceptibility of $Ba_{1-x}K_xTi_2Sb_2O$ ($x$ = 0, 0.04, 0.05, and 0.1) measured at 20 kOe.

### 3.4 Phase diagram of $Ba_{1-x}K_xTi_2Sb_2O$

On the basis of the above results the superconducting, magnetic and structural phase diagram of hole-doped $Ba_{1-x}K_xTi_2Sb_2O$ ($0 \leq x \leq 0.15$) is presented in Figure 5. Qualitatively, the resemblance to the phase diagrams of the other hole-doped systems $Ba_{1-x}Na_xTi_2Sb_2O$ and $BaTi_2(Sb_{1-x}Sn_x)_2O$ is apparent [19,27]. Namely, the CDW transition is gradually suppressed to lower temperatures by substitution with potassium, sodium and tin, respectively, and $T_a$ forms a downward concave curve in the region of higher doping level $x$ showing a saturation tendency. $T_c$ gradually increases exhibiting likewise a saturation tendency in the higher doped regime, which is supposed to be related to the robustness of the CDW phase [27], which co-exists with superconductivity.

Quantitatively, however, the phase diagram of $Ba_{1-x}K_xTi_2Sb_2O$ differs slightly from that of $Ba_{1-x}Na_xTi_2Sb_2O$, and significantly from the one of $BaTi_2(Sb_{1-x}Sn_x)_2O$. By doping with potassium instead of sodium, the CDW ordering temperature is suppressed slightly faster to lower temperatures, whereby the rise of $T_c$ correspondingly proceeds more quickly. The maximum $T_c$ of about 6.1 K is slightly higher than the one for $Ba_{1-x}Na_xTi_2Sb_2O$ featuring thus the highest $T_c$ in this family of layered titanium oxypnictides so far. Another difference is that for $Ba_{1-x}Na_xTi_2Sb_2O$ superconductivity is observed for doping levels of up to 33 %, whereas superconductivity in $Ba_{1-x}K_xTi_2Sb_2O$ is largely suppressed already at a doping level of 20 %. According to the data presented by Doan *et al.* [19], it is conceivable that superconductivity at sodium doping levels higher than 25 % may be filamentary due to phase separation. As for sodium doping levels higher than 30 - 35 % no homogeneous samples could be obtained, inhomogeneity might even start at lower doping levels concerning small volume fractions. Thus, it could be possible that for samples with a nominal Na doping level higher than 25 % small amounts of a 25 % or less doped phase coexists evoking the superconducting signal. This would also agree with the rather unusual fact that for higher doping levels $T_c$ remains constant, while the volume fraction decreases.

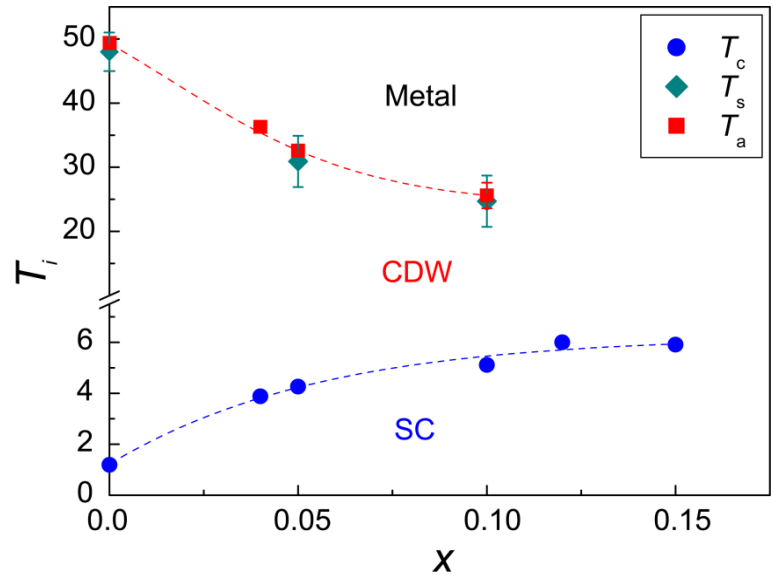

Figure 5: Phase diagram of $Ba_{1-x}K_xTi_2Sb_2O$ showing temperatures of the superconducting transitions ($T_c$), magnetic ($T_a$) and structural anomalies ($T_s$).

A quite remarkable result of our study is the fact that the phase diagrams of sodium- and potassium-doped $BaTi_2Sb_2O$ are so similar in spite of the significantly reverse volume effect (see Figure 2). Even though the unit cell volume slightly increases by potassium doping while it significantly decreases by sodium doping, neither the CDW transition temperature $T_a$ nor the superconducting critical temperatures $T_c$ are responsive to it. This is rather unusual, as CDW states are expected to depend on the volume via lattice hardening or softening under physical or chemical pressure as known from other CDW systems [12]. On the other hand, our results are reminiscent to the iron-arsenides $Ba_{1-x}K_xFe_2As_2$ and $Ba_{1-x}Na_xFe_2As_2$ [29,30], which likely exhibit very similar phase diagrams with respect to $T_c$ and $T_a$ in spite of quite different volume effects. From this one may conclude that the charge variation plays the dominant role in controlling the phase diagrams of $Ba_{1-x}K_xTi_2Sb_2O$ and $Ba_{1-x}Na_xTi_2Sb_2O$.

## 4    Conclusion

The complete solid solution $Ba_{1-x}K_xTi_2Sb_2O$ ($0 \leq x \leq 1$) was synthesized and characterized by X-ray powder diffraction and magnetic measurements. Potassium-doping increases the superconducting transition temperature from 1.2 K in the $BaTi_2Sb_2O$ parent compound up to 6.1 K ($x = 0.12$), which is the highest value in the family of $ATi_2Pn_2O$ oxypnictides. Superconductivity diminishes at doping levels higher than 0.15, where the superconducting phase fraction is already reduced. Anomalies in the magnetic susceptibility ascribed to a CDW transition shift to lower temperatures upon potassium doping, and are no longer discernible at higher doping levels. Thus superconductivity and CDW order co-exist up to $x \approx$ 0.15, and are both absent at higher potassium concentrations. The phase diagrams of $Ba_{1-x}K_xTi_2Sb_2O$ and $Ba_{1-x}Na_xTi_2Sb_2O$ are very similar, which is quite remarkable with respect to the inverse volume effect. Potassium doping increases the unit cell volume, while sodium doping reduces it. Obviously neither the CDW ordering nor the superconducting critical temperature significantly depend on the unit cell volumes. From this we conclude that the charge variation induced by hole-doping plays the dominant role, and probably determines the phase diagrams of $Ba_{1-x}K_xTi_2Sb_2O$ and $Ba_{1-x}Na_xTi_2Sb_2O$.

**Acknowledgement**

This work was financially supported by the German Research Foundation (DFG) within the priority program SPP1459, project Jo257/6-2).